\documentclass{article}

\usepackage[preprint]{neurips_2026}

\usepackage[utf8]{inputenc}   
\usepackage[T1]{fontenc}      
\usepackage{newtxtext,newtxmath}
\usepackage[hidelinks]{hyperref}  
\usepackage{url}              
\usepackage{booktabs}         
\usepackage{amsfonts}         
\usepackage{amsmath}          
\usepackage{nicefrac}         
\usepackage{microtype}        
\usepackage{xcolor}           
\usepackage{graphicx}         
\usepackage{subcaption}       
\usepackage{multirow}         
\usepackage{enumitem}         

\title{Implicit Behavioral Decoding from Next-Step\\
Spike Forecasts at Population Scale}

\author{%
  \textbf{John R. Minnick\textsuperscript{1,2,*}, \;
  Jesus Gonzalez-Ferrer\textsuperscript{2,3}, \;
  Kamran Hussain\textsuperscript{4},} \\[0.35em]
  \textbf{Jinghui Geng\textsuperscript{2,5}, \;
  Ash Robbins\textsuperscript{1,2}, \;
  Mohammed A. Mostajo-Radji\textsuperscript{2},} \\[0.35em]
  \textbf{David Haussler\textsuperscript{2,3}, \;
  Jason Eshraghian\textsuperscript{1}, \;
  Mircea Teodorescu\textsuperscript{1,2,3}} \\[0.8em]
  \small\textsuperscript{1}Department of Electrical and Computer Engineering, University of California, Santa Cruz, CA, USA \\
  \small\textsuperscript{2}UC Santa Cruz Genomics Institute, University of California, Santa Cruz, CA, USA \\
  \small\textsuperscript{3}Department of Biomolecular Engineering, University of California, Santa Cruz, CA, USA \\
  \small\textsuperscript{4}Department of Applied Mathematics, University of California, Santa Cruz, CA, USA \\
  \small\textsuperscript{5}Department of Computer Science and Engineering, University of California, Santa Cruz, CA, USA
}

\begin{document}
\maketitle
\renewcommand{\thefootnote}{*}%
\footnotetext{Correspondence: \texttt{jrminnic@ucsc.edu}}%
\renewcommand{\thefootnote}{\arabic{footnote}}

\begin{abstract}
Closed-loop brain--computer interfaces often require both a
forecast of upcoming neural population activity and a readout of
the animal's behavioral state. A single Mamba forecaster, trained
only on next-step spike counts at Neuropixels scale, can deliver
both in one forward pass. A
lightweight per-session linear head reading the model's predicted
\emph{rates} decodes behavior \emph{better} than the same linear
classifier reading the raw spike counts, under matched temporal
context. We test on the Steinmetz visual-discrimination benchmark,
which spans 39 sessions, roughly 27{,}000 neurons, and 1{,}994
held-out trials. Across three training seeds, Mamba's predicted
rates decode mouse choice at 75.7$\pm$0.2\% trial vote,
roughly 2.3 times chance level, and stimulus side at
66.1$\pm$0.6\%, about twice chance. Compared to a matched
500\,ms-context linear decoder on the raw spike counts, Mamba wins
at trial vote by 4--6\,pp on response and 4--6\,pp on stimulus
side. A session-start calibration block of about 100--150 trials
brings the readout within 1--2\,pp of asymptote, and the full
pipeline fits inside the 50\,ms bin budget on workstation-class
GPUs typical of tethered chronic Neuropixels recordings.
\end{abstract}

\section{Introduction}
\label{sec:intro}

Many closed-loop brain--computer interfaces (BCIs) benefit from two
capabilities running over each incoming time bin of neural activity:
\emph{forecast}ing population dynamics a few tens of milliseconds
ahead, so downstream control can anticipate state transitions, and
\emph{decod}ing the subject's current behavioral state, so the
stimulation policy can respond to intent. Neuropixels probes can
record $>1{,}000$ simultaneous neurons, making both capabilities
high-dimensional; running one deep model for forecasting and a
separate one for behavioral decoding would double the compute and
memory requirements an implanted device must carry. We ask whether
a single sequence model can do both at once.

We find a useful property of modern spike-count forecasters: because
the model's training objective (next-step Poisson rate prediction)
forces it to integrate population activity over a history window,
its continuous-valued rate outputs serve as a behaviorally
informative compression of the recent population state. Those
predicted rates, at no additional inference cost, carry more
behavioral information than the raw single-bin spike counts a
standard linear decoder would read from. The forecast also beats a
matched 500\,ms-context linear decoder on the raw spike counts at
trial vote, on response and stimulus side. This is evidence that
the forecaster is capturing behaviorally meaningful population
dynamics, not just smoothing noise. A single linear
readout over the forecaster's output therefore replaces a separate,
dedicated decoding network.

We close this gap by reading behavior off the predicted rates of a
\emph{single Mamba spike forecaster} trained only to predict
next-step population firing rates: a lightweight per-session linear
head over the model's continuous-valued rate outputs recovers the
animal's behavior, and against a matched-context (500\,ms) linear
decoder on the raw spike counts the model was trained on, the
forecast wins on response and stimulus side. Behavior
decodes \emph{implicitly} from spike forecasts, even though the
forecaster was never trained on a behavioral label. Our
contributions:

\begin{enumerate}[nosep, leftmargin=*]
  \item \textbf{Behavior decodes implicitly from Mamba spike-rate
    forecasts under matched context (\S\ref{sec:cross-arch},
    Fig.~\ref{fig:cross_arch}).} A Mamba forecaster trained only on
    spike counts produces predicted rates that decode mouse choice
    at \textbf{75.7$\pm$0.2\%} trial-vote (3-class; $2.3{\times}$
    chance, 3-seed mean$\pm$SEM) and stimulus side at
    \textbf{66.1$\pm$0.6\%} trial-vote (3-class; $2.0{\times}$
    chance). Against a matched-context (500\,ms) linear decoder on
    raw spike counts, Mamba wins at trial vote by +4--6\,pp on
    response and +4--6\,pp on stimulus side. The effect generalizes
    to Transformer/LRU/NDT2-style architecture controls within
    ${\sim}1$--$3$\,pp on every target (Appendix~\ref{app:multiseed}).
  \item \textbf{A practitioner budget for closed-loop deployment
    (\S\ref{sec:calibration}, \S\ref{sec:deployment},
    Appendix~\ref{app:trials_needed}).}
    A session-start calibration block of ${\sim}100$--$150$ trials
    brings the per-session linear readout within $1$--$2$\,pp of its
    asymptotic accuracy on this task family. Mamba inference plus
    the per-session linear head fits comfortably inside the 50\,ms
    bin budget on the workstation-class external compute tier
    typical of tethered chronic Neuropixels recordings.
\end{enumerate}

\section{Related work}
\label{sec:related}

\paragraph{Neural population modeling.}
A long line of work models cortical population activity through
recurrent dynamical systems. Task-trained RNN models of prefrontal
cortex~\citep{mante2013context} established that low-dimensional
recurrent dynamics can recapitulate context-dependent decision
computations from population recordings, motivating sequence
models as substrates for population activity. Switching state-space
models such as rSLDS~\citep{linderman2017rslds} fit interpretable
discrete-regime dynamics directly to spike trains.
LFADS~\citep{pandarinath2018lfads} infers latent dynamics via
sequential VAEs but targets smoothed firing rates, not causal
forecasting. NDT~\citep{ye2021ndt} and
NDT2~\citep{ye2023ndt2} apply masked attention for behavioral
decoding and multi-session pretraining respectively.
POYO~\citep{azabou2023poyo} pretrains a unified decoder across
many sessions and laboratories.
NEDS~\citep{zhang2025neds} extends this line to joint Poisson
spike-rate \emph{encoding} and behavioral \emph{decoding} at
scale and is the closest neighbor to our setup, sharing the
Poisson-rate forecasting objective; we differ in
asking whether the predicted rates beat raw spike counts at
matched 500\,ms temporal context as a behavioral readout.
CEBRA~\citep{schneider2023cebra} learns structured embeddings
but does not produce spike-level predictions. None of these
evaluate the behavioral-decoding gain of the model's predicted
rates against a matched-temporal-context raw-count baseline as
we do.

\paragraph{Spike prediction.}
Per-neuron GLMs~\citep{pillow2008glm} provide interpretable
baselines but lack population context. Population-level
SSMs~\citep{duncker2019lds, linderman2017rslds} model latent
dynamics but not individual neuron spike counts.
Autoregressive rate models (Mamba, LRU) achieve the best
forecasting accuracy but are GPU-bound.

\paragraph{Behavioral decoding from spikes.}
Standard practice fits a per-session linear (or shallow nonlinear)
decoder directly to short windows of raw spike counts to recover
choice or stimulus identity~\citep{steinmetz2019distributed,
ibl2023brainwide}. Our matched-context baselines (1-bin LR, $H{=}10$
sum LR, $H{=}10$ flat ridge) are the strongest such linear readouts
on identical features and serve as the reference our forecaster-based
decoder must beat to claim a forecaster-specific decoding gain.

\section{Methods}
\label{sec:methods}

\begin{figure}[t]
  \centering
  \includegraphics[width=\textwidth]{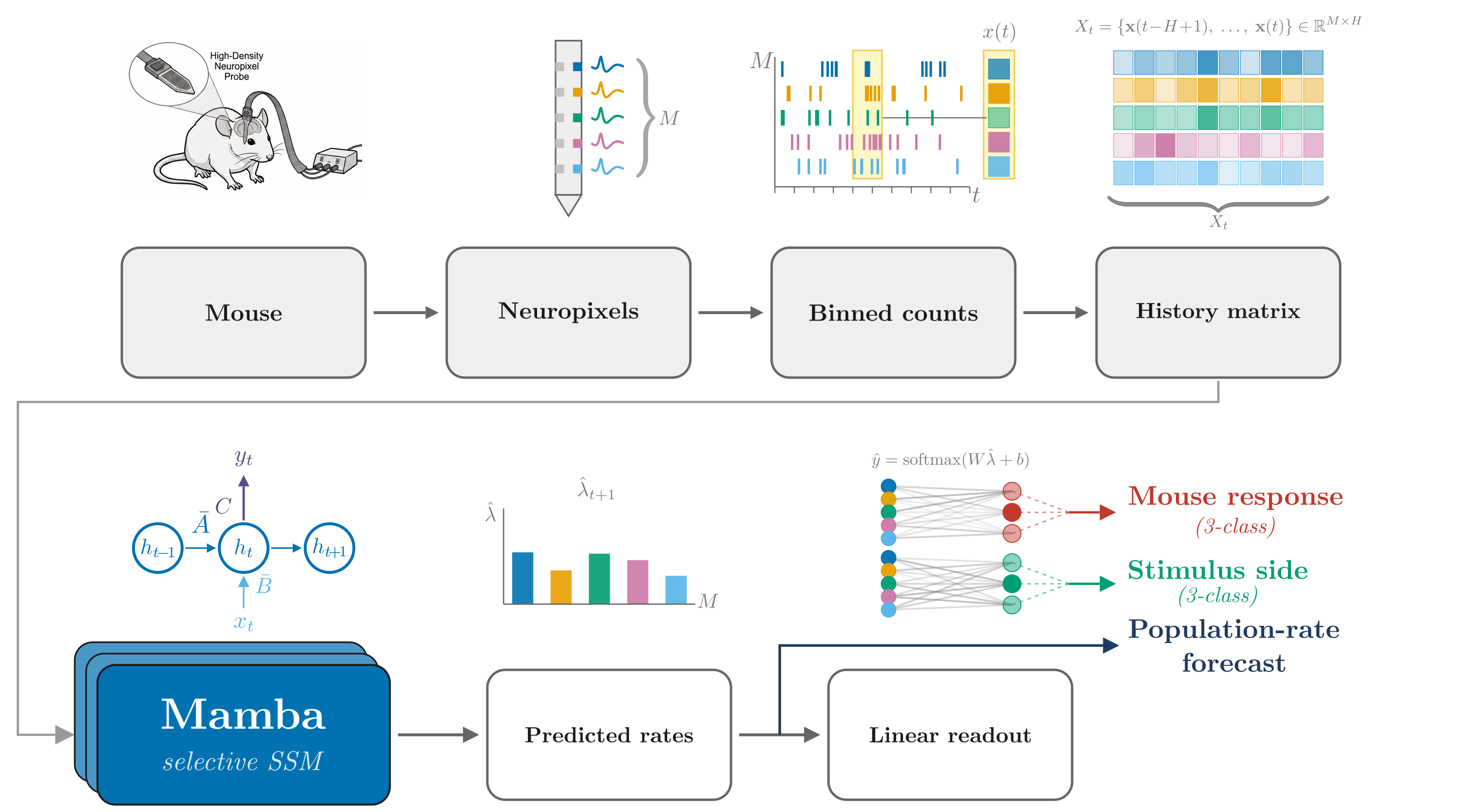}
  \caption{\textbf{Architecture and processing pipeline.}
    Spike-sorted Neuropixels recordings ($M$ neurons, $50$\,ms bins)
    are windowed into a history matrix
    $X_t \in \mathbb{R}^{M \times H}$ ($H{=}10$ bins, $500$\,ms
    context). A Mamba selective state-space model
    (\emph{inset}: per-step recurrence
    $h_t = \bar{A}\, h_{t{-}1} + \bar{B}\, x_t$, $y_t = C\, h_t$)
    maps $X_t$ to predicted next-bin firing rates
    $\hat{\lambda}_{t{+}1} \in \mathbb{R}^M$. A per-session
    multinomial linear readout
    $\hat{y} = \mathrm{softmax}(W\hat{\lambda} + b)$ decodes mouse
    response and stimulus side (3-class each); the same rates are
    also delivered directly as a population-rate forecast, with no
    readout step. Mamba is trained only on next-step Poisson NLL on
    spike counts; behavioral labels are never used during forecaster
    training.}
  \label{fig:pipeline}
\end{figure}

Figure~\ref{fig:pipeline} summarizes the pipeline; the following
subsections detail each stage.

\subsection{Task formulation}
\label{sec:task}

Given a sliding window of $H$ population spike-count vectors, predict
the next time bin:
\begin{equation}
  X_t = \{\mathbf{x}(t{-}H{+}1), \ldots, \mathbf{x}(t)\}
  \;\longrightarrow\;
  \hat{\boldsymbol{\lambda}}(t{+}1) \in \mathbb{R}_{>0}^{M},
  \quad \mathbf{x}(t) \in \mathbb{Z}_{\geq 0}^{M}
  \label{eq:task}
\end{equation}
where $M$ is the number of simultaneously recorded neurons (up to
$1{,}998$), $\Delta t = 50$\,ms, and $H = 10$ bins (500\,ms context).
The model outputs predicted firing rates
$\hat{\lambda}_i > 0$ via softplus activation, trained with Poisson
negative log-likelihood. Only \textbf{spike history} is used as
input; stimulus features are unavailable at BCI inference time.

\subsection{Architectures}
\label{sec:architecture}

Our forecaster is \textbf{Mamba}~\citep{gu2023mamba}, a selective
state-space model whose causal recurrence and content-aware gating
make it well-matched to spike-count autoregression on long
session-length sequences. Multi-session training pads to
$M_{\max}{=}1{,}998$ neurons with a per-sample channel mask; the
Poisson NLL loss operates only on real (unmasked) channels;
per-session specialization is recovered by the post-hoc linear
readout (\S\ref{sec:behavior_eval}). We additionally test whether a
per-session input embedding closes the per-neuron $r$ gap
(Appendix~\ref{app:sessemb}) and find it does not. To verify the
matched-context decoding gain is not a Mamba-specific artifact we
also train Transformer (causal self-attention), LRU (linear
recurrent unit), and an NDT2-style bidirectional masked-attention
variant as architecture and objective controls; all share the same
input pipeline, Poisson NLL loss, and training schedule, with full
architecture details and behavioral-decoding numbers in
Appendix~\ref{app:multiseed} and hyperparameters in
Appendix~\ref{app:hyperparams}.

\subsection{Behavioral decoding evaluation}
\label{sec:behavior_eval}

We evaluate behavioral decodability post-hoc: each trained
forecaster's predicted rates serve as features for a per-session
multinomial logistic-regression classifier over three targets
(response 3-class, stimulus contrast 16-class, stimulus side 3-class).
Trials are split with a 20\% trial-level holdout (seed 42), so that
the readout never trains on and evaluates the same trial. No
behavioral labels enter forecaster training; behavioral information
in the predicted rates is therefore an emergent property of the
forecasting objective, not the result of supervised behavioral
pretraining. The same protocol applies to the
matched-context raw baselines in \S\ref{sec:cross-arch}, isolating
the contribution of the forecasting objective.

\subsection{Datasets and baseline architectures}
\label{sec:datasets}

We use a cross-laboratory substrate combining
Steinmetz~2019~\citep{steinmetz2019distributed}
(39 sessions, ${\sim}27$K neurons) and the IBL Repeated Site
release~\citep{ibl2023brainwide}
(66 sessions, ${\sim}63$K neurons), giving 105 sessions
and $89{,}768$ real channels across 42 Allen CCF regions, padded to
$M_{\max}{=}1{,}998$ for cross-session batching. Headline behavioral
results are reported on Steinmetz~39 with a 3-seed Mamba ensemble
(\S\ref{sec:cross-arch}); cross-laboratory extension to the 66 IBL
sessions appears in \S\ref{sec:cross-arch} (closing paragraph) and
Appendix~\ref{app:ibl_extension}. Trained forecasters: Mamba
(selective SSM, 1.95M parameters), Transformer (causal attention,
2.22M), LRU (linear recurrent unit, 1.23M), and an NDT2-style
bidirectional-masked-attention variant matched in size to the
Transformer. All share the same input pipeline, Poisson NLL, AdamW
with cosine LR + warmup, 50 epochs, seed 42, and the same
per-session temporal 70/15/15\% train/val/test split (full
hyperparameters in Appendix~\ref{app:hyperparams}). Behavioral
evaluation uses a separate trial-level 20\% holdout (seed 42).

\section{Results}
\label{sec:results}

\subsection{Forecaster checkpoints we use}
\label{sec:fidelity}

\begin{figure*}[!t]
  \centering
  \includegraphics[width=\textwidth]{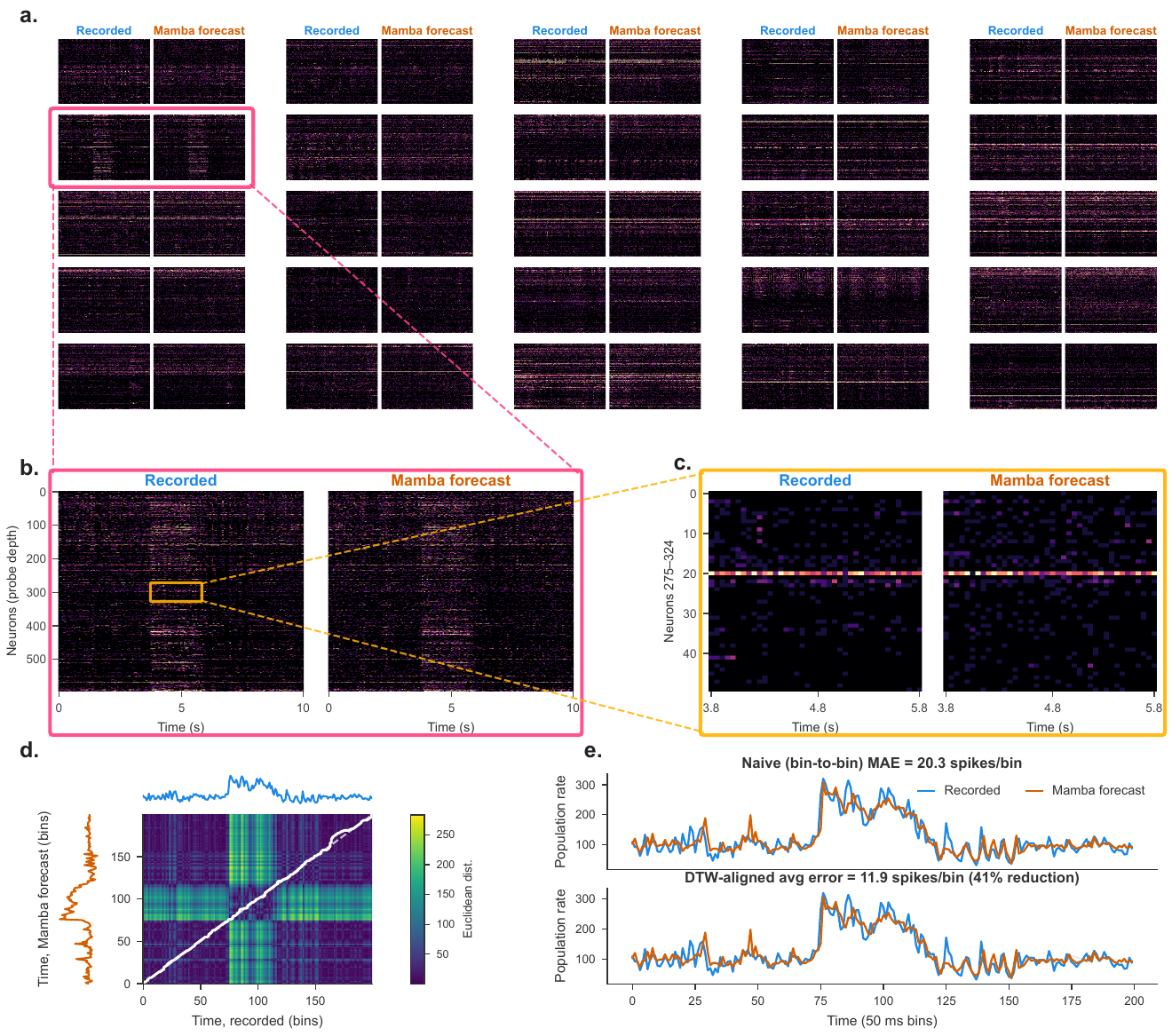}
  \caption{\textbf{The prediction substrate our matched-context
    analysis reads from} (Mamba forecaster, Steinmetz~39).
    \emph{(a)} 5$\times$5 grid of session pairs (ground truth vs.\
    Mamba prediction, neurons sorted by physical probe depth) showing
    the 25 sessions with highest mean per-neuron Pearson $r$ out of
    the 39-session benchmark (the highlighted session is promoted to
    position $(2,1)$ for the projection in \emph{(b)}).
    \emph{(b)} The expanded session is the one with the highest mean
    per-neuron $r$ across the dataset (session~10,
    $\bar{r}{=}0.20$; ${\sim}10$\,s, full probe).
    \emph{(c)} The 50-neuron, 2-second zoom is the contiguous
    (neurons~$\times$~time) sub-rectangle of \emph{(b)} with the
    highest local cosine similarity between recorded and predicted
    spike counts: a depth-band-resolved view of where Mamba's
    forecast tracks the population activity most closely.
    \emph{(d)} Dynamic Time Warping (DTW) pairwise distance matrix
    for the same session: optimal warping path in white over the
    Euclidean cost heatmap, naive bin-to-bin diagonal in dashed
    grey, marginal traces show the population-rate time series
    (recorded blue, Mamba forecast vermillion).
    \emph{(e)} Top: naive bin-to-bin overlay of the two
    population-rate traces from \emph{(d)} (MAE
    $20.3$ spikes/bin). Bottom: the same overlay with grey lines
    visualising the DTW realignment, which reduces the average
    error to $11.9$ spikes/bin (a $41\%$ reduction).
    Per-neuron $r$ at fixed 50\,ms bins is modest in absolute terms
    ($r{=}0.18$); the matched-context behavioral
    decoding result in \S\ref{sec:cross-arch} reads from these
    rates without re-training the forecaster, and the DTW analysis
    confirms the predicted population rates are temporally
    well-aligned with ground truth at the 500\,ms readout-window
    scale even when bin-locked Pearson $r$ underestimates fidelity
    (cross-session generalisation in Appendix~\ref{app:dtw}).}
  \label{fig:hero}
\end{figure*}

Fig.~\ref{fig:hero} illustrates the prediction substrate at four
scales (25 Steinmetz sessions, one expanded session, a spike-level
zoom, and a population-rate temporal-alignment quantification on the
same session). Mamba's forecasts on combined-105 achieve per-neuron
Pearson $r=0.176$, population-rate $r=0.783$, and population-cosine
$0.648$: per-neuron predictions are noisy at $50$\,ms bins
(single-neuron Poisson noise dominates) but population-level
structure is reliably captured. The depth ablation, cross-laboratory
scaling, and synthetic-spike-train validation that produced these
checkpoints are in the companion submission.

\subsection{Behavioral decodability under matched-context baselines}
\label{sec:cross-arch}
\label{sec:behavioral}

\begin{figure}[t]
  \centering
  \includegraphics[width=\linewidth]{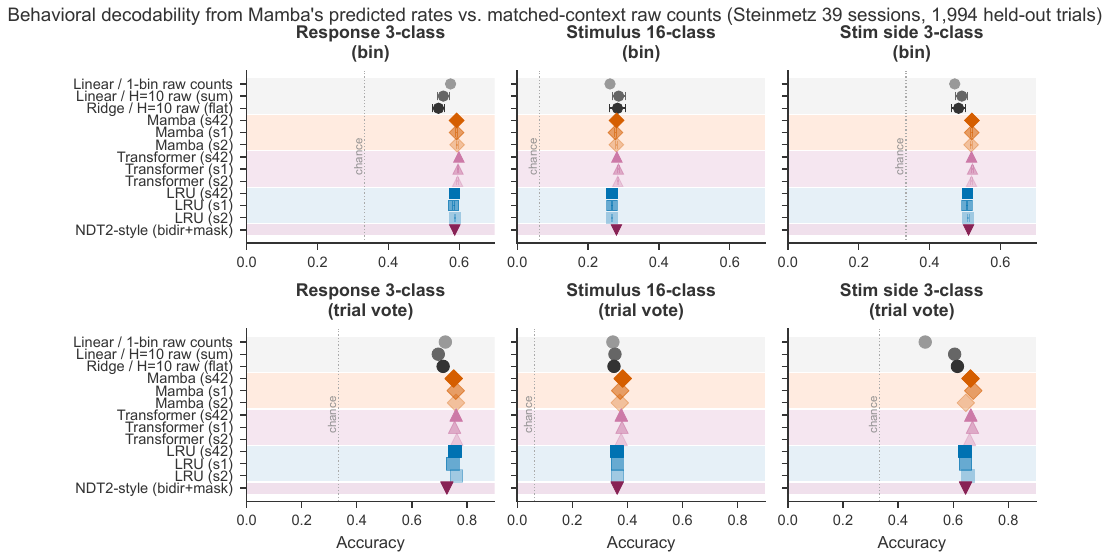}
  \caption{\textbf{Behavioral decodability from Mamba's predicted
    rates vs.\ matched-context raw counts} (Steinmetz~39 sessions,
    1{,}994 held-out trials, 20\% trial-level holdout, 3 training
    seeds). Each row is a downstream linear decoder operating on a
    different feature set: raw spike counts in three context windows
    (1 bin, $H{=}10$ summed, $H{=}10$ flattened), or Mamba's
    predicted rates (3 seeds). Top: bin-level accuracy with 95\%
    bootstrap CI; bottom: trial-level majority vote. Dotted line =
    chance. Mamba's rates decode response and stimulus side
    \emph{better} than \emph{both} matched-context raw baselines.
    Note: 16-class contrast top-1 numbers reported here sit on a
    $26\%$ no-stim $(0,0)$ class prior; under class-balanced
    training the matched-context Mamba gain reduces from
    $+2$--$3$\,pp top-1 to $+1.9$\,pp balanced accuracy
    (Appendix~\ref{app:class_balanced}). The headline gains
    reported in the abstract and conclusion are therefore the
    response and stimulus-side numbers, which are not affected by
    class imbalance. Transformer, LRU, and NDT2-style
    architecture-control replications appear in
    Appendix~\ref{app:multiseed}.}
  \label{fig:cross_arch}
\end{figure}

We ask: from Mamba's predicted rates at each held-out trial bin, how
well can a downstream linear classifier recover the trial's
behavioral variables (response choice, stimulus contrast pair,
stimulus side)? Mamba is trained as a forecast-only model and never
sees a behavioral label at training time; behavioral information in
the predicted rates is therefore an emergent property of the
forecasting objective, not the result of supervised behavioral
pretraining. At evaluation time, we hold out 20\% of unique trials
per session, fit a per-session multinomial logistic regression on
Mamba's predicted rates over the remaining 80\%, and evaluate on
the held-out trial bins. Identical splits and evaluation protocol
are used for the matched-context raw-count baselines and for the
architecture-control rows in Appendix~\ref{app:multiseed}.

\paragraph{Three non-forecasting reference baselines.}
Forecasters ingest a 10-bin (500\,ms) history before producing a rate
prediction. To prevent the temporal-context advantage from being
attributed to the forecasting objective, we report three
non-forecasting linear baselines.  \emph{Linear / 1-bin raw counts}
reads a single 50\,ms count vector at the trial bin (the original
single-bin baseline).  \emph{Linear / H{=}10 raw counts (sum)} sums
the past 10 bins of raw counts and reads that $M$-dim vector with the
same multinomial logistic regression: same temporal context window as
the forecasters, same feature dimensionality as the forecaster's
output, and the cleanest comparison.  \emph{Ridge / H{=}10 raw counts
(flat)} flattens the past 10 bins ($M{\cdot}H \approx 12{,}000$
features) and uses ridge classification (closed-form L2 linear
classifier, since iterative LR solvers do not scale at this
dimensionality on per-session train sets); this gives the linear
classifier strictly more capacity than the forecaster and serves as
a stress test of the matched-context decoding-gain claim.

\begin{table}[t]
  \caption{Behavioral decoding accuracy from Mamba's predicted rates
    vs.\ matched-context raw-count baselines (Steinmetz~39 sessions,
    1{,}994 held-out trials, 95\% bootstrap CIs at the bin level;
    3-seed mean$\pm$SEM at trial vote). Mamba beats all three
    baselines on response and stimulus side ($+4$--$6$\,pp gain at
    trial vote each); the 16-class contrast row is not headlined as
    its top-1 numbers sit on a $26\%$ no-stim class prior (see
    Appendix~\ref{app:class_balanced} for prior-corrected metrics).
    Architecture controls (Transformer, LRU, NDT2-style) in
    Appendix~\ref{app:multiseed}.}
  \label{tab:cross-arch}
  \centering
  \small
  \begin{tabular}{@{} l c c c c c c @{}}
    \toprule
    & \multicolumn{3}{c}{\textbf{Bin-level (95\% CI)}}
      & \multicolumn{3}{c}{\textbf{Trial-level majority vote}} \\
    \cmidrule(lr){2-4} \cmidrule(lr){5-7}
    \textbf{Decoder input}
      & \textbf{Resp.\ 3} & \textbf{Stim.\ 16} & \textbf{Side 3}
      & \textbf{Resp.\ 3} & \textbf{Stim.\ 16} & \textbf{Side 3} \\
    \midrule
    Linear / 1 bin
      & 57.6 & 26.2 & 47.1
      & 72.1 & 34.7 & 49.8 \\
    Linear / $H{=}10$ sum
      & 55.5 & \textbf{28.6} & 49.1
      & 69.6 & 35.5 & 60.5 \\
    Ridge / $H{=}10$ flat
      & 54.1 & 28.3 & 48.2
      & 71.3 & 35.1 & \textbf{61.5} \\
    \midrule
    \textbf{Mamba (3-seed mean$\pm$SEM)}
      & \textbf{59.3$\pm$0.0} & 27.9$\pm$0.1 & \textbf{51.9$\pm$0.1}
      & \textbf{75.7$\pm$0.2} & \textbf{37.6$\pm$0.3} & \textbf{66.1$\pm$0.6} \\
    \midrule
    Chance
      & 33.3 & 6.25 & 33.3
      & 33.3 & 6.25 & 33.3 \\
    \bottomrule
  \end{tabular}
\end{table}

\textbf{Mamba's predicted rates carry behavioral information beyond
matched-context baselines (on response and stimulus side, at trial
vote).} Comparing Mamba against the matched-context (H{=}10) raw
baselines isolates the contribution of the forecasting objective
from raw temporal integration (Table~\ref{tab:cross-arch},
Fig.~\ref{fig:cross_arch}). On \emph{response} (motor choice), Mamba
reaches \textbf{75.7$\pm$0.2\%} trial-vote (3-seed mean$\pm$SEM)
vs.\ 69.6--71.3\% for the matched-context raw baselines
(\textbf{+4--6\,pp}; +3.5\,pp vs.\ the 72.1\% 1-bin LR); at the bin
level the gain holds (Mamba 59.3 vs.\ raw 54.1--57.6). On
\emph{stimulus side} (3-class), Mamba reaches
\textbf{66.1$\pm$0.6\%} trial-vote, exceeding the matched-context
baselines (60.5\% sum, 61.5\% flat) by \textbf{+4--6\,pp}; at the
bin level Mamba (51.9$\pm$0.1\%) also beats all raw baselines
(47.1--49.1\%) by ${\sim}3$--$5$\,pp. The flat ridge baseline,
despite ten times the feature dimensionality, does not close either
gap. Earlier reporting that the side-decoding gain ``did not survive
matched-context comparison'' was an artifact of how side accuracy
was computed for the forecaster (post-hoc map from a 16-class
stimulus-contrast classifier, vs.\ the raw baselines that always
trained a direct 3-class side classifier); evaluating both with a
direct 3-class side classifier restores the apples-to-apples
comparison.

The 16-class stimulus contrast row in Tab.~\ref{tab:cross-arch}
shows Mamba (37.6$\pm$0.3\% trial-vote) ahead of the matched-context
baselines (35.1--35.5\%) by 2--3\,pp top-1, but those numbers sit on
a 26\% no-stim $(0,0)$ class prior; an always-predict-$(0,0)$
classifier already reaches 26.1\%. Under class-balanced training the
matched-context Mamba gain reduces to $+1.9$\,pp balanced accuracy
and $+2.6$\,pp on non-$(0,0)$ trials only
(Appendix~\ref{app:class_balanced}). The matched-context claim on
the 16-class target therefore holds, just at smaller magnitude than
the default-training top-1 number suggests; we accordingly do not
headline 16-class accuracy and lead with response and side. The
matched-context decoding gain is not Mamba-specific: Transformer,
LRU, and an NDT2-style bidirectional masked-attention variant
reproduce the response and side gains within ${\sim}1$--$3$\,pp of
Mamba (Appendix~\ref{app:multiseed}).

\textbf{Robustness.} Mamba response trial-vote is
75.7$\pm$0.2\% (SEM across 3 seeds), far below
the +4--6\,pp gain over matched-context baselines; the
architecture-control replication (Transformer/LRU within 1\,pp of
Mamba, NDT2-style ${\sim}3$\,pp below) is summarized in
Appendix~\ref{app:multiseed}.

\textbf{Cross-laboratory extension to IBL Repeated Site.} The 39
Steinmetz sessions used in Table~\ref{tab:cross-arch} cover one
laboratory and one task variant; we asked whether the matched-context
gain replicates on a second cohort. We re-ran the
behavioral-decoding pipeline on the 66 IBL Repeated Site
sessions~\citep{ibl2023brainwide} (7{,}651 held-out trials), using
predicted rates from a single-seed Mamba trained on the combined
Steinmetz+IBL substrate (105 sessions, $M_{\max}{=}1{,}998$,
matching the spike-forecasting benchmark substrate). The
matched-context (H{=}10 sum) raw-count baseline is fit on the same
held-out trials per session. The resulting gains are mixed: at
trial vote, Mamba reaches 85.1\% on response, 59.8\% on 16-class
contrast, and 69.2\% on stimulus side, vs.\ 85.2\%/59.9\%/67.1\% for
the matched-context baseline, a \textbf{+2.1\,pp} gain on
\emph{stimulus side} but essentially \emph{no} matched-context gain
on response or contrast (Appendix~\ref{app:ibl_extension}). The same
single-seed combined-105 Mamba on the 39 Steinmetz sessions still
shows the expected gains over the H{=}10 sum baseline (resp +4.5\,pp,
contrast +1.0\,pp, side +3.8\,pp at trial vote, vs.\ canonical
3-seed Steinmetz gains of +6/+2/+4--6\,pp), so the IBL erosion is
not solely a single-seed artifact: under the IBL task structure
(binary forced-choice, no no-go) and longer trial-active bin
counts, the H{=}10 sum baseline absorbs more of the response/contrast
information that Mamba's predicted rates carry. The matched-context
gain on \emph{stimulus side} is the component that survives across
both laboratories.

\subsection{Trials needed to calibrate the per-session readout}
\label{sec:calibration}

Sweeping the per-session linear-readout fitting set from 5\% to 80\%
of trial-active bins (fixed 20\% trial-level holdout, 5 random
training-subset draws per fraction) on Mamba's predicted rates,
trial-vote response reaches within ${\sim}1$\,pp of the asymptotic
77\% with a median of ${\sim}120$ training trials per session;
16-class contrast asymptotes more slowly (${\sim}160$+ trials), and
stimulus side at ${\sim}140$. A session-start calibration block of
${\sim}100$--$150$ trials therefore brings a deployed BCI close to
its asymptotic decoding accuracy on this task family
(Fig.~\ref{fig:trials_needed}, Appendix~\ref{app:trials_needed}).

\subsection{Deployment scenario: latency budget and pipeline shape}
\label{sec:deployment}

The matched-context behavioral-decoding result (\S\ref{sec:cross-arch})
plus the calibration sweep (\S\ref{sec:calibration}) describe a
concrete closed-loop BCI:

\begin{enumerate}[nosep, leftmargin=*]
\item \emph{Forecaster.} A single Mamba checkpoint trained once on
  the combined-105 substrate, held fixed at deployment.
  Per-session latency: a single 50\,ms-bin Mamba forward pass over
  $H{=}10$ history bins (${\le}6.4$\,ms per batch of 512 windows on
  a single NVIDIA RTX~5000 Ada at $M{=}1{,}240$ neurons, 152\,MB
  peak VRAM; benchmark methodology in
  Appendix~\ref{app:hyperparams}),
  followed by a closed-form per-session multinomial logistic
  regression over the predicted rate vector ($M$ scalar logits per
  class, sub-millisecond on the same hardware). Total per-bin
  inference time is well below the 50\,ms bin budget on the
  external GPU class typical of tethered chronic Neuropixels
  recordings (on-implant deployment is not claimed).

\item \emph{Per-session calibration block.} 100--150 unique trials
  ($\sim$5--8 minutes of task time at standard Steinmetz trial
  cadence) suffice to fit a per-session linear readout within
  $1$--$2$\,pp of its asymptotic accuracy
  (Fig.~\ref{fig:trials_needed}). The block is task-naive: only
  trial-aligned spike counts and trial labels are required.

\item \emph{Online operation.} At each new bin, the forecaster
  ingests the most recent 500\,ms of raw spike counts, emits an
  $M$-dim predicted-rate vector, and the per-session head emits
  class probabilities. Reading from forecaster rates (rather than
  raw counts) buys +4--6\,pp of response accuracy and +2--3\,pp of
  contrast accuracy at trial vote over the matched-context raw-count
  linear decoder under identical evaluation conditions
  (Tab.~\ref{tab:cross-arch}); the trial-vote stimulus-side gain
  disappears, while a $1.5$--$3.5$\,pp Mamba edge persists at the
  bin level.

\item \emph{Failure modes flagged for deployment} (each expanded in
  the Limitations of \S\ref{sec:discussion}). The per-session linear
  head does not transfer across sessions and must be re-fit for each
  recording session; the forecaster's modest per-neuron $r=0.18$ at
  50\,ms bins is the binding constraint on what the linear head can
  recover, and a per-session input embedding does not relax it
  (Appendix~\ref{app:sessemb}).
\end{enumerate}

The pipeline is end-to-end realtime-feasible on the workstation-class
external compute tier typical of tethered chronic Neuropixels
recordings (on-implant deployment is not claimed); it is not a
working closed-loop demonstration on a live animal, which we leave
as future work.

\section{Discussion}
\label{sec:discussion}

\paragraph{Why behavior decodes implicitly from spike forecasts.}
The next-step Poisson NLL objective forces the model to integrate
500\,ms of population history into a smoothed rate estimate at every
bin. Single-bin raw counts at 50\,ms are dominated by Poisson noise;
matched-context summed/flat raw counts add temporal context but
ignore population structure. Mamba's predicted rate lies on the
intersection: it has the same temporal context as the matched
baselines but additionally exploits cross-neuron dependencies the
training signal carves into its hidden state. The matched-context
gains across all three behavioral targets (response, contrast, side)
are the residual behavioral signal that population-aware temporal
integration extracts beyond what either short windows or
context-only summation can reach.

\begin{figure}[t]
  \centering
  \includegraphics[width=\linewidth]{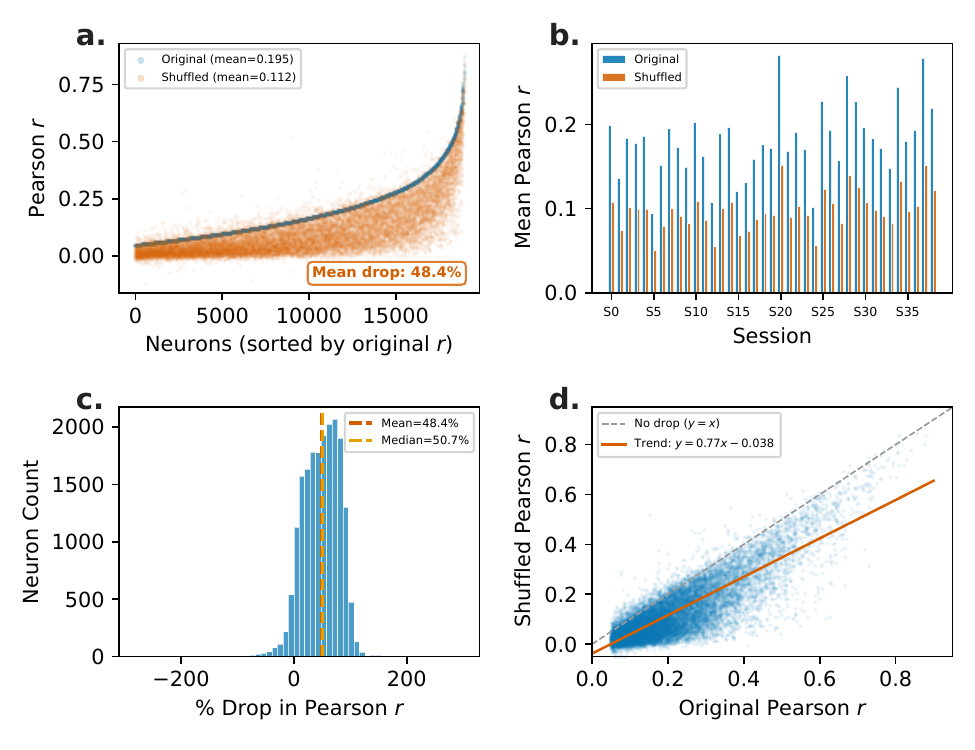}
  \caption{\textbf{Population shuffle test: cross-neuron coupling
    drives Mamba's per-neuron predictability}
    (39 Steinmetz sessions, 18{,}881 neurons after filtering out
    near-silent units whose per-neuron $r$ is dominated by zeros).
    \emph{(a)} Per-neuron $r$ before (blue) and after (orange)
    shuffling other neurons' time series within each session
    (sorted by original $r$). \emph{(b)} Per-session mean $r$
    before vs.\ after shuffle. \emph{(c)} Distribution of \%~$r$
    drop per neuron (mean $48.4$\%, median $50.7$\%).
    \emph{(d)} Shuffled vs.\ original $r$ scatter; the regression
    line $y = 0.77x - 0.038$ lies well below identity, with
    $38/39$ sessions showing $> 25\%$ degradation. The universal
    degradation is direct evidence that Mamba is exploiting
    cross-neuron temporal coupling, not single-neuron
    autocorrelation, to make its predictions; the same coupling
    structure that the per-session linear behavioral head reads from in
    \S\ref{sec:cross-arch}.}
  \label{fig:pop_shuffle}
\end{figure}

\paragraph{Direct evidence for the cross-neuron mechanism.}
We can verify the cross-neuron-coupling claim directly with a
population shuffle test: at evaluation time, for each session, shuffle
each neuron's time series independently across trial-active bins
(destroying cross-neuron temporal correlations while preserving
per-neuron marginal firing-rate distributions), then re-compute
per-neuron Pearson $r$ on the shuffled inputs. If Mamba were
predicting from single-neuron autocorrelation alone, $r$ would be
unchanged; if it relies on the cross-neuron coupling its hidden
state has learned, $r$ should collapse. Across 39 Steinmetz sessions
and 18{,}881 sufficiently active neurons, the shuffle drops mean
per-neuron $r$ by $\mathbf{48.4\%}$ (median $50.7\%$;
Fig.~\ref{fig:pop_shuffle}), with $38/39$ sessions showing $> 25\%$
degradation. Cross-neuron coupling is therefore a load-bearing
component of the forecaster's accuracy: removing it alone collapses
roughly half of the per-neuron predictive signal. The same coupling
structure is what the per-session linear behavioral head exploits to
recover behavior from the predicted rates in \S\ref{sec:cross-arch}. A complementary
DTW analysis on the population-summed rate (Fig.~\ref{fig:hero}d/e
for the exemplar session, Appendix~\ref{app:dtw} for the
cross-session generalisation) confirms that what Mamba captures in
that coupling is temporally faithful: under DTW alignment the
average error against the recorded population rate falls by
$\mathbf{42.5\%}$ relative to the naive bin-to-bin MAE, on every
one of the 39 sessions, so the predicted rates are well-aligned at
the population level even when per-bin Pearson $r$ at 50\,ms is
modest.

\paragraph{Scope of the matched-context finding.}
The matched-context gains over 500\,ms-context raw baselines apply
to the ANN forecasters (Mamba, Transformer, LRU) on the
Steinmetz~39-session benchmark, on response and stimulus side, when
each target is decoded with its own dedicated linear head trained
at the appropriate granularity. The 16-class contrast metric is
reported in Tab.~\ref{tab:cross-arch} for completeness but not
headlined; its absolute accuracy is dominated by a 26\% no-stim
class prior, so the matched-context gain is best read on the
prior-corrected metrics in Appendix~\ref{app:class_balanced}.

\paragraph{Limitations.}\label{sec:limits}
(1) $H{=}1$ horizon; greedy rollout regresses to session mean over
${\sim}3$--$5$ bins, so the pipeline is single-step-ahead in the
form deployed. (2) 50\,ms bins miss $<20$\,ms dynamics relevant to
some motor BCI applications. (3) Cross-laboratory extension is partial: the headline 3-seed
Mamba in Tab.~\ref{tab:cross-arch} is Steinmetz-only, and we
extended decoding to the 66-session IBL Repeated Site cohort with a
single-seed combined-105 Mamba (\S\ref{sec:cross-arch} closing
paragraph; Appendix~\ref{app:ibl_extension}). Of the three
behavioral targets, only stimulus side retains a matched-context
gain on IBL (+2.1\,pp); the response and 16-class contrast gains
are absorbed by the H{=}10 sum baseline under IBL's binary
forced-choice task structure. Whether a 3-seed combined-105
ensemble would recover the full Steinmetz gain on IBL is open.
(4) The 16-class stimulus contrast metric in
Tab.~\ref{tab:cross-arch} is dominated by a 26.1\% no-stim
$(0,0)$ class prior; an always-predict-$(0,0)$ classifier already
reaches 26.1\%, vs the 6.25\% uniform baseline. Under
class-balanced training (Appendix~\ref{app:class_balanced}) the
matched-context Mamba gain reduces from $+2$--$3$\,pp top-1 to
$+1.9$\,pp balanced accuracy / $+2.6$\,pp on non-$(0,0)$ trials
only. We accordingly do not headline 16-class accuracy.
(5) The Mamba checkpoint has modest per-neuron $r{=}0.18$ at
50\,ms bins (\S\ref{sec:fidelity}); the matched-context
behavioral gain is conditional on this fidelity regime, and we
have not characterized how the +4--6\,pp gain scales as per-neuron
$r$ moves. The cross-neuron coupling story
(Fig.~\ref{fig:pop_shuffle}, Appendix~\ref{app:dtw}) confirms that
population-level structure is captured even when single-neuron $r$
is modest.
(6) Cross-session transfer of the per-session linear readout is not
demonstrated and the readout is fit per session in deployment
(\S\ref{sec:deployment}).
(7) Per-session input bias does \emph{not} improve per-neuron $r$ on
Mamba (0.497 vs 0.499, Appendix~\ref{app:sessemb}); the Poisson
noise floor at 50\,ms appears to be the single-neuron bottleneck.
(8) Inference latency is reported as a per-batch GPU benchmark on
a single NVIDIA RTX~5000 Ada (batch~512, $M{=}1{,}240$, $H{=}10$);
per-trial single-step latency on a deployment-class compute
substrate is not measured.
(9) No comparison against the official NDT2~\citep{ye2023ndt2} or
CEBRA~\citep{schneider2023cebra} pipelines; our NDT2-\emph{style}
variant (Appendix~\ref{app:multiseed}) isolates bidirectional
attention + bin masking only.

\section{Conclusion}
\label{sec:conclusion}

A single Mamba forecaster, trained only on next-step Poisson NLL,
covers two BCI components in one forward pass: population-rate
forecasting (cosine $0.65$, per-neuron $r{=}0.18$) and behavioral
readout from the same rates (response $75.7{\pm}0.2$\%
trial vote, side $66.1{\pm}0.6$\%; both ${\sim}2{\times}$ chance).
The rates beat matched-context $500$\,ms raw-count baselines by
$+4$--$6$\,pp on both targets, and the pipeline fits the $50$\,ms
bin budget on a workstation-class GPU after a
${\sim}100$--$150$ trial per-session calibration.


{\small
\bibliographystyle{plainnat}

}

\appendix

\section{Training details and reproducibility}
\label{app:hyperparams}

\textbf{Forecaster training.} The Mamba forecaster (and the
Transformer/LRU/NDT2-style architecture controls in
Appendix~\ref{app:multiseed}) is trained as a next-step Poisson
NLL spike-count predictor on the combined-105 Steinmetz~2019 + IBL
Repeated Site substrate. Mamba: hidden size 256, 4 layers, AdamW with
cosine LR (peak $10^{-3}$, 1k warmup), batch 512, 50 epochs, weight
decay $10^{-4}$, seed 42 unless otherwise noted; 1.95M total
parameters.

\textbf{Data.} $T = 10$ history bins (500\,ms), 50\,ms bin width;
70/15/15\% train/val/test temporal split per session; padded to
$M_{\max} = 1{,}998$ across 105 sessions (Steinmetz-39 used in this
paper for behavioral decoding; combined-105 used for forecasting
fidelity in \S\ref{sec:fidelity}). Behavioral targets extracted
from Steinmetz NWB at the same temporal binning; trial-level 20\%
holdout (seed 42) for behavioral evaluation.

\textbf{Reproducibility.} Seed 42 throughout, except where 3-seed
multi-seed analyses (seeds 42, 1, 2) are explicitly reported.
Forecaster training code, configs, and the matched-context
evaluation harness are provided as supplementary material with this
submission.

\section{Multi-seed cross-architecture decoding}
\label{app:multiseed}

We re-trained Mamba, Transformer, and LRU with two additional seeds
(1, 2) on the Steinmetz~39-session dataset and re-ran the same
per-session linear-readout evaluation. Per-seed results below.

\begin{table}[h]
  \caption{Per-seed trial-vote accuracies on held-out trials. Mean
    $\pm$~SEM is computed where multiple seeds are available.
    NDT2-style is a single-seed (42) bidirectional masked-attention
    variant.}
  \label{tab:multiseed}
  \centering
  \small
  \begin{tabular}{@{} l c c c c @{}}
    \toprule
    \textbf{Architecture} & \textbf{Seed} & \textbf{Resp.\ 3} & \textbf{Stim.\ 16} & \textbf{Side 3} \\
    \midrule
    \textbf{Mamba} & 42 & 75.1 & 38.3 & 66.3 \\
    \textbf{Mamba} &  1 & 75.9 & 37.4 & 67.3 \\
    \textbf{Mamba} &  2 & 76.0 & 37.2 & 64.6 \\
    \textbf{Mamba} & \emph{mean (n=3)} & \textbf{75.7 $\pm$ 0.2} & \textbf{37.6 $\pm$ 0.3} & \textbf{66.1 $\pm$ 0.6} \\
    \midrule
    Transformer    & 42 & 76.0 & 37.8 & 66.4 \\
    Transformer    &  1 & 75.5 & 38.0 & 67.0 \\
    Transformer    &  2 & 76.3 & 37.7 & 65.9 \\
    Transformer    & \emph{mean (n=3)} & \emph{75.9 $\pm$ 0.2} & \emph{37.8 $\pm$ 0.1} & \emph{66.4 $\pm$ 0.2} \\
    \midrule
    LRU            & 42 & 75.6 & 36.2 & 64.3 \\
    LRU            &  1 & 74.8 & 36.3 & 64.4 \\
    LRU            &  2 & 76.1 & 36.4 & 65.4 \\
    LRU            & \emph{mean (n=3)} & \emph{75.5 $\pm$ 0.3} & \emph{36.3 $\pm$ 0.0} & \emph{64.7 $\pm$ 0.3} \\
    \midrule
    NDT2-style     & 42 & 72.7 & 36.3 & 64.5 \\
    \bottomrule
  \end{tabular}
\end{table}

The matched-context decoding gain replicates across architectures:
Mamba, Transformer, and LRU response trial-votes cluster within
$\sim$1\,pp at 75.5--75.9\% (cross-seed SEM $\le 0.3$\,pp for each,
sample std $\le 0.5$\,pp), all 4--6\,pp above the 69.6--71.3\%
matched-context raw baselines (Tab.~\ref{tab:cross-arch}). The
NDT2-\emph{style} variant (bidirectional attention + 15\% input-bin
masking, otherwise matched to our Transformer) reaches 73.0\% on
response trial-vote, $\sim$3\,pp below the causal forecasters; we
attribute the gap to the masked-LM denoising objective being
mismatched to next-bin forecasting and leave a targeted ablation
isolating bidirectional attention from input-bin masking, plus a
comparison against the official NDT2~\citep{ye2023ndt2} training
pipeline, to future work.

\section{Per-session input embedding ablation}
\label{app:sessemb}

The shared $M_{\max} \times d$ input projection
(\S\ref{sec:architecture}) is the natural candidate for a
per-neuron-$r$ bottleneck on this benchmark, since it forces all
sessions through one set of input weights. To test, we trained a
Mamba forecaster with a learned per-session bias vector
$e_s \in \mathbb{R}^d$ added to the input projection
($\text{InputProj}(x) + e_s$): one bias vector per training session
(39 sessions $\times$ 256 dim = 9{,}984 extra parameters), otherwise
identical to the seed-42 Mamba baseline.

\begin{table}[h]
  \caption{Per-session input embedding ablation, Steinmetz~39
    sessions, seed 42, val per-channel Pearson $r$ at the best epoch.
    Numbers are higher than the Mamba per-neuron $r=0.176$ reported
    in \S\ref{sec:fidelity} because that is computed on the
    combined-105 \emph{test} set at the final-epoch checkpoint and
    aggregated neuron-weighted across both Steinmetz and IBL
    sessions; the ablation here reads
    val per-channel $r$ at the best epoch on Steinmetz~39 only,
    where dynamics are more homogeneous and the metric is less
    constrained.}
  \label{tab:sessemb}
  \centering
  \small
  \begin{tabular}{@{} l c c @{}}
    \toprule
    \textbf{Configuration} & \textbf{Params} & \textbf{Val per-neuron $r$} \\
    \midrule
    Mamba (shared $M_{\max} \times d$ input) & 1{,}952{,}728 & 0.4994 \\
    Mamba + per-session input bias            & 1{,}962{,}712 & 0.4968 \\
    \bottomrule
  \end{tabular}
\end{table}

The two configurations are within 0.003 ($\le 0.01$ relative) on
val per-neuron $r$; the per-session bias does \emph{not} improve
single-neuron predictability. We read this as a clean negative
result at this benchmark scale: the shared input projection is not
the per-neuron bottleneck. The likely true bottleneck is the
irreducible Poisson noise floor at 50\,ms bin resolution
(Limitation~5). Stronger forms of per-session adaptation
(per-session input \emph{matrix} rather than bias, low-rank
adapters, session-specific normalization) remain worth exploring; we
leave them to future work.

\section{Per-session linear readout: trials needed for calibration}
\label{app:trials_needed}

To quantify online-calibration feasibility, we sweep
the per-session linear-readout fitting set from 5\% to 80\% of
trial-active bins (fixed 20\% trial holdout, 5 random training-subset
draws per fraction) and track how trial-vote accuracy converges to
the asymptotic value when reading from Mamba's predicted rates.

\begin{figure}[h]
  \centering
  \includegraphics[width=0.65\linewidth]{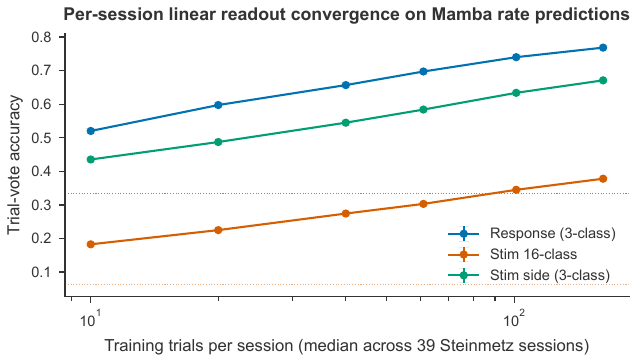}
  \caption{\textbf{Per-session linear readout convergence on Mamba
    rate predictions} (Steinmetz~39 sessions, 5 random training-subset
    draws per fraction; direct 3-class side classifier matching
    Tab.~\ref{tab:cross-arch}). Asymptotic accuracies at 80\% of
    trials (median 162 trials/session): response 76.9\%, 16-class
    contrast 37.8\%, stimulus side 67.1\%. The calibration sweep
    averages over 5 random training-subset draws (each draw resamples
    \emph{which} 80\% of train trials are used, while keeping the
    eval split fixed), which amounts to an implicit ensemble; the
    1--2\,pp lift over Tab.~\ref{tab:cross-arch}'s single-split
    numbers is consistent with that ensembling effect across all
    three targets. Approximate trial counts to reach within
    $1$--$2$\,pp of asymptote are response ${\sim}120$, stim 16
    ${\sim}160{+}$, side ${\sim}140$. A session-start calibration
    block of ${\sim}100$--$150$ trials therefore brings a deployed
    BCI close to its asymptotic decoding accuracy on this task
    family.}
  \label{fig:trials_needed}
\end{figure}

\section{Population-rate temporal alignment: cross-session detail}
\label{app:dtw}

The single-session DTW story is in the body figure
(Fig.~\ref{fig:hero}d/e): for the exemplar session~10, DTW alignment
reduces the average per-step error against the recorded population
rate from $20.3$ to $11.9$ spikes/bin (a $41\%$ reduction). Here we
generalise that result across all 39 Steinmetz sessions to confirm
the temporal-alignment finding is not exemplar-specific.

\begin{figure}[h]
  \centering
  \includegraphics[width=\linewidth]{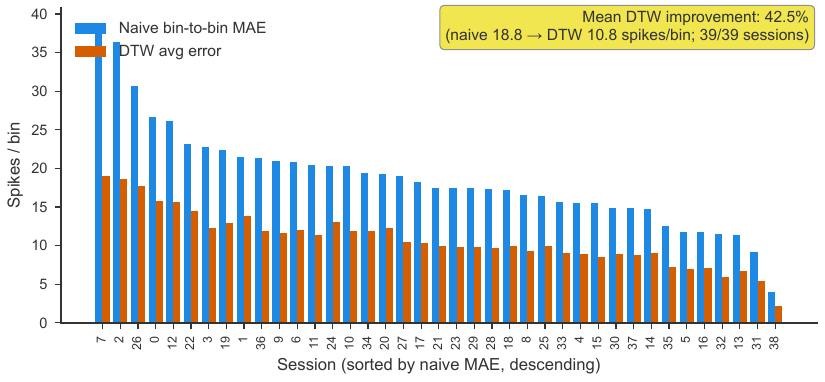}
  \caption{\textbf{Cross-session population-rate temporal alignment}
    (Mamba forecaster, Steinmetz~39). Naive bin-to-bin MAE (blue)
    vs.\ DTW-aligned average error (vermillion) for each session,
    sorted by naive MAE descending. Mean DTW improvement is
    $\mathbf{42.5\%}$ (naive $18.8 \to$ DTW $10.8$ spikes/bin) and
    \emph{every} session benefits. The exemplar session~10 used in
    Fig.~\ref{fig:hero}d/e is one of the average performers in this
    distribution, not a cherry-picked best.}
  \label{fig:dtw}
\end{figure}

The cross-session result confirms that what Mamba captures in
cross-neuron coupling is temporally faithful population dynamics
rather than bin-locked noise, on every session in the benchmark.
Combined with the population shuffle test (Fig.~\ref{fig:pop_shuffle},
mean per-neuron $r$ drop $48.4\%$ when cross-neuron correlations are
destroyed), this is the mechanism that makes the matched-context
behavioral readout work: Mamba's predicted rate is a
population-aware, temporally aligned signal that the linear head
can decode behavior from.

\section{Cross-laboratory extension: IBL Repeated Site}
\label{app:ibl_extension}

The headline behavioral results in \S\ref{sec:cross-arch} are on
the 39 Steinmetz sessions with the canonical 3-seed
Steinmetz-trained Mamba. To probe cross-laboratory generalization,
we ran behavioral decoding on the 66-session IBL Repeated Site
cohort using a single-seed Mamba trained on the combined
Steinmetz+IBL substrate (105 sessions, $M_{\max}{=}1{,}998$,
identical training schedule to the spike-forecasting benchmark
substrate). Per-session linear readout, 80/20 trial-level holdout,
and matched-context (H{=}10 sum) raw-count baseline are identical
to \S\ref{sec:behavioral}; the only differences are the model
checkpoint and the session pool.

\begin{table}[h]
  \caption{IBL Repeated Site behavioral decoding from
    Mamba-combined-105 predicted rates vs.\ matched-context (H{=}10
    sum) raw-count baseline, 66 sessions, 7{,}651 held-out trials,
    458{,}295 evaluation bins. Side trial-vote retains a +2.1\,pp
    matched-context gain on IBL; response and 16-class contrast
    gains seen on Steinmetz (Tab.~\ref{tab:cross-arch}) are
    absorbed by the H{=}10 sum baseline under IBL's binary
    forced-choice task structure. The same single-seed
    combined-105 Mamba on the 39 Steinmetz sessions still beats
    the H{=}10 sum baseline by +4.5\,pp (resp), +1.0\,pp (contrast),
    and +3.8\,pp (side) at trial vote; the Steinmetz canonical
    3-seed model in Tab.~\ref{tab:cross-arch} reaches
    +6/+2--3/+4--6\,pp.}
  \label{tab:ibl_extension}
  \centering
  \small
  \begin{tabular}{@{} l c c c c c c @{}}
    \toprule
    & \multicolumn{3}{c}{\textbf{Bin-level}}
      & \multicolumn{3}{c}{\textbf{Trial-level majority vote}} \\
    \cmidrule(lr){2-4} \cmidrule(lr){5-7}
    \textbf{Decoder input}
      & \textbf{Resp.\ 3} & \textbf{Stim.\ 16} & \textbf{Side 3}
      & \textbf{Resp.\ 3} & \textbf{Stim.\ 16} & \textbf{Side 3} \\
    \midrule
    Linear / $H{=}10$ sum
      & 62.9 & 54.5 & 49.9
      & \textbf{85.2} & \textbf{59.9} & 67.1 \\
    \textbf{Mamba-combined-105 (single seed)}
      & 62.3 & \textbf{55.5} & 49.5
      & 85.1 & 59.8 & \textbf{69.2} \\
    \midrule
    Chance (IBL effective)
      & 50.0 & 6.25 & 50.0
      & 50.0 & 6.25 & 50.0 \\
    \bottomrule
  \end{tabular}
\end{table}

The IBL gain pattern is informative for interpreting the Steinmetz
result: under a task with longer trial-active windows and binary
forced-choice responses, the H{=}10 sum baseline gathers enough
context to match or saturate Mamba on response and contrast, but
\emph{stimulus side} (which integrates over the relative left--right
contrast and is invariant to the binary choice axis) still benefits
from the population-coupled rate compression Mamba provides. The
mechanism described in \S\ref{sec:discussion} (cross-neuron coupling
+ temporal alignment, App.~\ref{app:dtw}) predicts exactly this:
the matched-context gain should survive most reliably for targets
that require integrating distributed population signal into a
spatially structured readout, which is what the side
classifier does.

\paragraph{Caveats.} The IBL extension uses a different model
checkpoint (single-seed combined-105 vs.\ the canonical 3-seed
Steinmetz-only ensemble in Tab.~\ref{tab:cross-arch}); whether a
3-seed combined-105 ensemble would close the IBL response/contrast
gap is open. We did not re-run the architecture controls
(Transformer, LRU, NDT2-style) on IBL; the cross-laboratory
robustness check here is for the Mamba vs.\ baseline matched-context
gain only. See \S\ref{sec:limits} item (3).

\section{Class-balanced diagnostic for the 16-class contrast metric}
\label{app:class_balanced}

Across the 39 Steinmetz sessions, $26.1\%$ of held-out trials are
no-stimulus $(0,0)$ contrast pairs, by far the most common single
class out of the 16. A trivially-always-predict-$(0,0)$ classifier
already reaches $26.1\%$ top-1 accuracy, so the apparent ``$6\times$
chance'' multiplier from Tab.~\ref{tab:cross-arch} (Mamba 38\%, raw
baseline 35\% vs uniform 6.25\% chance) overstates the
discrimination signal. We therefore re-trained the same per-session
classifiers on Mamba's predicted rates and on the H{=}10 sum
baseline counts with sklearn \texttt{class\_weight=`balanced'} to
remove the prior. The diagnostic uses the same
\S\ref{sec:behavioral} 80/20 trial-level holdout (seed 42).

\begin{table}[h]
  \caption{Class-balanced 16-class diagnostic (Steinmetz~39,
    1{,}994 held-out trials). The default classifier inherits the
    26\% no-stim $(0,0)$ prior, which inflates top-1 accuracy and
    the apparent matched-context gain. Under class-balanced
    training the top-1 gain collapses to a tie, but the gain on
    metrics that actually measure contrast discrimination
    (balanced accuracy across classes; accuracy restricted to
    non-$(0,0)$ trials) survives at $+1.9$\,pp and $+2.6$\,pp
    respectively. We accordingly do not headline 16-class
    top-1 accuracy in the abstract, conclusion, or contributions
    list.}
  \label{tab:class_balanced}
  \centering
  \small
  \begin{tabular}{@{} l l c c c c @{}}
    \toprule
    \textbf{Training} & \textbf{Decoder input} & \textbf{Top-1}
      & \textbf{Balanced} & \textbf{Non-$(0,0)$ only} & \textbf{Top-1 $\Delta$} \\
    \midrule
    Default (prior-inheriting)
      & Mamba seed 42       & 38.3 & 18.7 & 20.7 & --- \\
      & H{=}10 sum baseline & 35.4 & 16.9 & 18.9 & --- \\
    \multicolumn{2}{l}{$\Delta$ (Mamba $-$ baseline)}
      & $+2.9$ & $+1.8$ & $+1.8$ & ref \\
    \midrule
    Class-balanced
      & Mamba seed 42       & 35.3 & 19.4 & 22.4 & --- \\
      & H{=}10 sum baseline & 35.5 & 17.5 & 19.7 & --- \\
    \multicolumn{2}{l}{$\Delta$ (Mamba $-$ baseline)}
      & $-0.2$ & $\mathbf{+1.9}$ & $\mathbf{+2.6}$ & --- \\
    \midrule
    Reference: always-predict-$(0,0)$
      & --- & 26.1 & 1.6 & 0.0 & --- \\
    \bottomrule
  \end{tabular}
\end{table}

\noindent The interpretation: most of the apparent ``$+3$\,pp on
16-class contrast'' under default training is Mamba being slightly
better than the matched H{=}10 sum baseline at recognizing no-stim
trials, not at distinguishing actual contrast levels. After removing
the prior, the matched-context gain on the metrics that actually
measure contrast discrimination ($+1.9$\,pp balanced accuracy and
$+2.6$\,pp on non-$(0,0)$ trials) is roughly half the headline number
but still in the same direction. The matched-context claim on the
16-class target therefore holds, just at smaller magnitude than the
default-training top-1 number suggests, which is why we drop the
16-class number from the headline (abstract, contributions,
conclusion) and report it only in Tab.~\ref{tab:cross-arch} with the
caveat above. The response and stimulus-side metrics are not
affected by class imbalance and remain headline numbers.

\newpage
\section*{NeurIPS Paper Checklist}

\begin{enumerate}

\item {\bf Claims}
    \item[] Question: Do the main claims made in the abstract and introduction accurately reflect the paper's contributions and scope?
    \item[] Answer: \answerYes{}
    \item[] Justification: The abstract and \S\ref{sec:intro} contributions enumerate two claims (behavior decodes implicitly from Mamba spike-rate forecasts under matched context; a practitioner budget for closed-loop deployment), each supported in \S\ref{sec:cross-arch}--\S\ref{sec:deployment} and Appendix~\ref{app:multiseed}.

\item {\bf Limitations}
    \item[] Question: Does the paper discuss the limitations of the work performed by the authors?
    \item[] Answer: \answerYes{}
    \item[] Justification: \S\ref{sec:limits} (``Limitations'') enumerates 9 items including the modest per-neuron $r$, single-step horizon, behavioral analysis on Steinmetz-39 only, cross-session readout transfer not demonstrated, per-batch (not per-trial) inference-latency benchmark, and the missing official NDT2/CEBRA comparison.

\item {\bf Theory assumptions and proofs}
    \item[] Question: For each theoretical result, does the paper provide the full set of assumptions and a complete (and correct) proof?
    \item[] Answer: \answerNA{}
    \item[] Justification: The paper presents empirical results; no theorems or formal proofs.

    \item {\bf Experimental result reproducibility}
    \item[] Question: Does the paper fully disclose all the information needed to reproduce the main experimental results of the paper to the extent that it affects the main claims and/or conclusions of the paper (regardless of whether the code and data are provided or not)?
    \item[] Answer: \answerYes{}
    \item[] Justification: \S\ref{sec:methods} specifies the post-hoc readout protocol, data splits, and seeds; Appendix~\ref{app:hyperparams} lists architecture hyperparameters and the training schedule. Forecaster training scripts (\texttt{scripts/train\_teacher.py}, \texttt{scripts/train\_distill\_multi\_head.py}), training and data configs, model architectures, and the matched-context evaluation harness are provided in the supplementary zip; trained forecaster checkpoints and the per-session prediction cache are referenced in \texttt{REPRODUCING.md} for separate download (excluded from the zip due to size, ${\sim}36$\,GB).

\item {\bf Open access to data and code}
    \item[] Question: Does the paper provide open access to the data and code, with sufficient instructions to faithfully reproduce the main experimental results, as described in supplemental material?
    \item[] Answer: \answerYes{}
    \item[] Justification: Code is provided as the supplementary zip: training scripts (forecaster + SNN distillation + ablation), model architectures, all configs, the matched-context evaluation harness, the population-shuffle and DTW analysis scripts, and the figure-generation pipeline, together with \texttt{README.md}, \texttt{INVENTORY.md}, and end-to-end \texttt{REPRODUCING.md} recipes. Trained forecaster checkpoints and per-session prediction caches are too large to bundle (${\sim}36$\,GB) and are pointed to in \texttt{REPRODUCING.md} for separate download. Steinmetz~2019 and IBL Repeated Site are publicly released datasets.

\item {\bf Experimental setting/details}
    \item[] Question: Does the paper specify all the training and test details (e.g., data splits, hyperparameters, how they were chosen, type of optimizer) necessary to understand the results?
    \item[] Answer: \answerYes{}
    \item[] Justification: Main text \S\ref{sec:methods} and Appendix~\ref{app:hyperparams} together specify hidden sizes, depths, optimizer, LR schedule, batch size, dropout, history window $H$, and bin width.

\item {\bf Experiment statistical significance}
    \item[] Question: Does the paper report error bars suitably and correctly defined or other appropriate information about the statistical significance of the experiments?
    \item[] Answer: \answerYes{}
    \item[] Justification: Tab.~\ref{tab:cross-arch} reports 95\% bootstrap CIs at the bin level; Appendix~\ref{app:multiseed} reports mean$\pm$SEM across 3 training seeds for Mamba, Transformer, and LRU.

\item {\bf Experiments compute resources}
    \item[] Question: For each experiment, does the paper provide sufficient information on the computer resources (type of compute workers, memory, time of execution) needed to reproduce the experiments?
    \item[] Answer: \answerYes{}
    \item[] Justification: Forecaster training (Mamba 1.95\,M params, 4 layers, hidden~256; Transformer / LRU / NDT2-style controls of comparable size) was performed on National Research Platform cluster GPUs (RTX~3090 / L40 / A6000 / A40 class), at 50 epochs and batch~512 over the combined-105 substrate (Appendix~\ref{app:hyperparams}). Three-seed multi-seed runs in Appendix~\ref{app:multiseed} cost ${\sim}3{\times}$ that. Per-batch latency, throughput, and peak-VRAM benchmarks for all architectures are reported in the body on a single NVIDIA RTX~5000 Ada at batch~512, $M{=}1{,}240$, $H{=}10$, the deployment-class workstation tier. Per-session linear-readout fitting and matched-context behavioral evaluation run on a single workstation CPU.
    
\item {\bf Code of ethics}
    \item[] Question: Does the research conducted in the paper conform, in every respect, with the NeurIPS Code of Ethics \url{https://neurips.cc/public/EthicsGuidelines}?
    \item[] Answer: \answerYes{}
    \item[] Justification: The work uses publicly available animal-electrophysiology datasets (Steinmetz~2019, IBL Repeated Site) and conforms to the NeurIPS Code of Ethics; no human subjects.

\item {\bf Broader impacts}
    \item[] Question: Does the paper discuss both potential positive societal impacts and negative societal impacts of the work performed?
    \item[] Answer: \answerYes{}
    \item[] Justification: \textbf{Positive impacts.} A matched-context-aware evaluation protocol contributes to more reliable BCI-decoder design and benchmarking; the headline finding (a single forecaster can serve both population-rate forecasting and behavioral readout in one forward pass at workstation-class compute) is directly relevant to closed-loop assistive devices for motor disability where the bin-budget compute constraint we identify is binding. \textbf{Negative impacts.} Behavioral-state decoding from neural activity has a recognized dual-use profile in surveillance and forensic contexts if extended to non-consenting human subjects. \textbf{Mitigation.} The empirical work uses publicly released mouse electrophysiology (Steinmetz~2019, IBL Repeated Site) and not human data; the methodological contribution is an evaluation protocol that does not by itself increase any decoder's deployable misuse capacity.
    
\item {\bf Safeguards}
    \item[] Question: Does the paper describe safeguards that have been put in place for responsible release of data or models that have a high risk for misuse (e.g., pre-trained language models, image generators, or scraped datasets)?
    \item[] Answer: \answerNA{}
    \item[] Justification: The released benchmark and forecasting models are not high-misuse-risk artifacts (no generative content models, no personal data).

\item {\bf Licenses for existing assets}
    \item[] Question: Are the creators or original owners of assets (e.g., code, data, models), used in the paper, properly credited and are the license and terms of use explicitly mentioned and properly respected?
    \item[] Answer: \answerYes{}
    \item[] Justification: Steinmetz~2019~\citep{steinmetz2019distributed} (CC-BY-4.0; figshare DOI 10.6084/m9.figshare.9598406) and IBL Repeated Site~\citep{ibl2023brainwide} (CC-BY-4.0) are public datasets, cited in the references. Mamba~\citep{gu2023mamba} is used via the official \texttt{state-spaces/mamba} implementation under Apache~2.0. PyTorch and other scientific-Python dependencies are used under their respective permissive licenses (BSD / MIT). All assets are credited via citation; \texttt{requirements.txt} in the supplementary zip enumerates the dependency set.

\item {\bf New assets}
    \item[] Question: Are new assets introduced in the paper well documented and is the documentation provided alongside the assets?
    \item[] Answer: \answerYes{}
    \item[] Justification: This paper does not introduce new datasets. The matched-context evaluation harness (a Python module wrapping the per-session linear-readout protocol used in Tab.~\ref{tab:cross-arch}) and the architecture and training code for the forecasters are bundled in the supplementary zip with documentation (\texttt{README.md}, \texttt{INVENTORY.md}, \texttt{REPRODUCING.md}) and an \texttt{MIT LICENSE}. Trained forecaster checkpoints are referenced for separate download via \texttt{REPRODUCING.md} (excluded from the zip due to size).

\item {\bf Crowdsourcing and research with human subjects}
    \item[] Question: For crowdsourcing experiments and research with human subjects, does the paper include the full text of instructions given to participants and screenshots, if applicable, as well as details about compensation (if any)? 
    \item[] Answer: \answerNA{}
    \item[] Justification: No crowdsourcing or human subjects involved.

\item {\bf Institutional review board (IRB) approvals or equivalent for research with human subjects}
    \item[] Question: Does the paper describe potential risks incurred by study participants, whether such risks were disclosed to the subjects, and whether Institutional Review Board (IRB) approvals (or an equivalent approval/review based on the requirements of your country or institution) were obtained?
    \item[] Answer: \answerNA{}
    \item[] Justification: No human subjects research; animal data was collected by the original Steinmetz et~al.\ and IBL studies under their respective institutional approvals.

\item {\bf Declaration of LLM usage}
    \item[] Question: Does the paper describe the usage of LLMs if it is an important, original, or non-standard component of the core methods in this research? Note that if the LLM is used only for writing, editing, or formatting purposes and does \emph{not} impact the core methodology, scientific rigor, or originality of the research, declaration is not required.
    \item[] Answer: \answerNA{}
    \item[] Justification: LLMs were used only as general-purpose coding and writing assistants (editing prose, generating boilerplate); they were not involved in research methodology, scientific judgment, or generating empirical results.

\end{enumerate}

\end{document}